\documentstyle[12pt,dina4]{article}
\newcommand{\rme}{{\rm e}}
\newcommand{\be}{\begin{equation}}
\newcommand{\ee}{\end{equation}}
\newcommand{\bee}{\begin{eqnarray}}
\newcommand{\eee}{\end{eqnarray}}

\begin{document}

\title{Pion-nucleus reactions in a microscopic transport
model\footnote{Supported by
BMFT and GSI Darmstadt}}
\author{A. Engel\thanks{Part of the dissertation of A. Engel},
W. Cassing, U. Mosel,
 M. Sch\"afer, Gy. Wolf\thanks{present address:
GSI Darmstadt, on leave from CRIP, Budapest}
\\Institut f\"ur Theoretische Physik, Universit\"at Giessen\\
D--35392 Giessen, Germany}

\maketitle

\begin{abstract}

We analyse pion--nucleus reactions in a microscopic transport model
of the BUU type, which propagates nucleons, pions, deltas
and $N(1440)$-resonances explicitly in space and time. In particular
we examine pion absorption and  inelastic scattering
cross sections for pion kinetic
energies $T_\pi =85-315MeV$ and  various target masses. In general,
the mass--dependence of the experimental data is well described for energies
up to the $\Delta$--resonance ($\approx 160 MeV$)
 while the absorption cross sections are somewhat
overestimated for the higher energies.
In addition we study the possible dynamical effects
 of delta-- and pion--potentials in the medium on various observables
as well as alternative models for the in-medium $\Delta$-width.

\end{abstract}

\sloppy

\newpage

\section{INTRODUCTION}

Pion production in nucleus-nucleus reactions has been proposed in the early
80's
as an observable to test the nuclear equation of state
\cite{stoe78,sto82,sto86}.
Since then, and also for
more general questions about the properties of hadronic matter
at high density,
a lot of experimental and theoretical studies on pion production in heavy-ion
collisions have been performed \cite{sto82,sto86,har85,har87,wal87}.
Nowadays it is widely accepted that pion production is not the most suitable
process for studying the nuclear equation of state (EOS), especially in
inclusive experiments, since the pions, which are created in the hot
 and dense reaction zone, are absorbed
and reemitted several times before they leave the system such that the final
pions emerge from densities zones $\rho \approx 0.5\rho_0$ \cite{Wol93}.
Nevertheless, pion production is still an important field of present nuclear
research  due to the pion's important role in the overall reaction
dynamics and the equilibration of the system.

The most elaborate theoretical approaches for the description of pion
 production are microscopical kinetic
models which include the propagation of pions and nucleon--resonances as
well as their mutual interactions \cite{mos91,cas90p}.
Basically there are three different microscopical realizations:
the Intra Nuclear Cascade (INC)\cite{cug81,cug82,cah83,cug88},
 Quantum Molecular Dynamics (QMD) \cite{sor89,bas93},
and the Boltzman--Uehling--Uhlenbeck
model (BUU)\cite{wol90,wol92,mol85,stoe86,li91}.
In the INC model the nucleus--nucleus collision is simulated as the
sum of all individual nucleon--nucleon collisions without taking into
account self consistent mean--field potentials and Pauli blocking for the
collisions.
The QMD follows the same scheme as the INC, but takes into account the
Pauli blocking
in the collisions and a nucleus potential which is
calculated as the sum of all two--body potentials.
In this paper we will use the BUU--model  \cite{Wol93}
which is quite successful in
describing the experimental data on pion production in proton--nucleus
as well as nucleus--nucleus collisions. Since the pion spectra and pion
yields turn out to be quite insensitive to details of the treatment
of the $\pi N\Delta$--dynamics in the BUU codes for heavy--ion reactions,
 we
 study in this paper explicitly
 pion--nucleus reactions. We note that similar studies have already
been performed with the INC model several years
ago \cite{har73,gin78,fra82,cug88}.

{}From the pion physics point of view there has been a vivid interest in
describing pion--nucleus reactions, because pion--nucleus reactions
are expected to provide   information on the properties of the
strong interaction and excited states of the nucleon in the medium.
Two major aspects are important
 for the understanding of pion--nucleus reactions:
first, nuclear structure effects are important when looking
for exclusive processes, i.e. detecting also the final state of
the nucleus, and, second, the reaction mechanism itself which determines also
inclusive reactions.
 Especially the question of the absorption mechanism is
widely discussed \cite{ash86,wey90,ing93,sal88} in connection
with multi--nucleon
contributions and the properties of the
delta resonance in the nuclear medium.

The most successful model up to date
to describe inclusive pion--nucleus data is
the model from Salcedo et al. \cite{sal88}. In a first step
these authors calculate microscopically
density--dependent probabilities for pion quasielastic scattering and
pion absorption in the nucleus, also taking into account three--body
processes. With
these probabilities they perform a pion cascade calculation using
 nuclear density profiles without, however,
  propagating the nucleons (or $\Delta$'s)
 explicitly.
We will  compare our results with their calculations.

In this paper we have the twofold goal to test the validity of the
pion and resonance dynamics of our BUU model and to investigate
the reaction mechanism in pion--nucleus reactions. In this respect
 we compare our calculations with total and differential
cross sections for inclusive measurements in the pion energy
region  $T_\pi=85MeV - 315MeV$ for pion absorptive and inelastic processes.
This energy regime is of particular interest since here the delta
is predominantly excited in the nucleus.
To address the above mentioned questions, we proceed in the following way:
in section 2 we review the basic inputs of the BUU model, especially
those which are relevant for pion--nucleus dynamics. In section 3 we
discuss the results of our model in comparison with various experimental
data, compare with previous calculations and study the
 dependence of observables
on the input of our model. In section 4 we give a summary.

\section{THE EXTENDED BUU MODEL}
\subsection{Basic equations}
\label{basics}

The dynamical description of hadron-nucleus or
nucleus-nucleus reactions is based
on the equation of motion for
the time evolution of the nucleon one-body phase-space distribution
$f(\vec r,\vec p;t)$

\be
\label{BUUeq}
\frac{\partial f(\vec r,\vec p_1;t)}{\partial t}
 + \frac{\vec p_1}{m} \nabla_{\vec r}
f(\vec r, \vec p_1;t)
-  \nabla_{\vec r}U(\vec r) \nabla_{\vec p_1}
f(\vec r,\vec p_1;t)  = I[f(\vec r,\vec p_1;t)]
\ee
with
\bee
\label{kol1}
I[f(\vec r,\vec p_1;t)] = \frac{4}{(2\pi )^3} \int d^3p_2 d^3p_3 \int
d\Omega_4 v_{12}
\frac{d\sigma}{d\Omega}
&& \nonumber \\
\delta^3(\vec p_1 +\vec p_2-\vec p_3-\vec p_4)
\times (f_3f_4\bar f_1 \bar f_2
 - f_1f_2\bar f_3 \bar f_4)     & &
\eee
where $U(\vec r)$ is the nucleon mean--field potential, $d\sigma /d\Omega$ is
the differential nucleon--nucleon cross section,
$\bar f_i = 1-f(\vec r,\vec p_i;t)$ the Pauli blocking  factors and $v_{12}$
the relative velocity between nucleon 1 and 2. Eq.\ (\ref{BUUeq}) is
called the Boltzmann--Uehling--Uhlenbeck equation (also known as
Vlasov--Uehling--Uhlenbeck, Boltzmann--Nordheim or Landau--Vlasov equation).
Its static homogeneus solution is the Fermi--Dirac distribution function.
Details of the derivation of eq.\ (\ref{BUUeq}) can be found in
ref. \cite{ber88,cas88}. Physically the lhs of eq.\ (\ref{BUUeq}) represents
a Vlasov equation for a gas of nucleons moving
independently in the mean field
U. The rhs of  eq.\ (\ref{BUUeq}) is the two--body collision integral $I[f]$
which
describes changes of the phase-space distribution due to nucleon--nucleon
collisions incorporating Pauli-blocking for the final nucleon states.

Above the pion threshold inelastic processes become more and more
important in nucleon--nucleon collisions such that pions, $\Delta$'s and
even  higher resonances
have to be included in the model. Therefore, in \cite{wol90,wol92}
 we have extended the BUU model
to coupled transport equations for
 pions, etas, {\mbox{$\Delta$--,}} $N(1440)$-- and $N(1535)$--resonance
 distribution
 functions \cite{wol90,wol92} (see also \cite{wan91}). Since for
the present study only nucleons,
deltas, $N(1440)$'s and pions play a role, we
discard the explicit propagation of etas and higher resonances.

 Schematically  one can write down the coupled equations used in our model
in the following way:
 \bee
 \label{BUUex}
 Df_N&=&  I^N_{NN}  + I^N_{N\Delta} + I^N_{NN^*} +I^N_{\Delta\rightarrow N\pi}
     + I^N_{N^*\rightarrow N\pi} \nonumber \\
 Df_\Delta &=& I^\Delta_{N\Delta} +I^\Delta_{\Delta\rightarrow N\pi}  \\
 Df_\pi &=& I^\pi_{\Delta\rightarrow N\pi} +
 I^\pi_{N^*\rightarrow N\pi} \nonumber
 \\
 Df_{N^*} &=& I^{N^*}_{N^*\rightarrow N\pi} + I^{N^*}_{N^*N} \nonumber
 \eee
 where $Df$ always denotes the differential operator for the
 Vlasov equation and the collision term on the rhs of eq.\ (\ref{BUUex})
 incoporates the following processes:
 \be
 \label{collproc}
 \begin{array}{lc}
 I^N_{N\Delta},I^\Delta_{N\Delta} & N\Delta \leftrightarrow N\Delta \,\,
 {\rm and} \,\, \Delta N \leftrightarrow NN \\[2mm]
 I^N_{NN^*} I^{N^*}_{NN^*} & NN^*\leftrightarrow NN^* \, \,
  {\rm and} \, \, NN^*\leftrightarrow NN \\[2mm]
  I^N_{\Delta\rightarrow N\pi}, I^\pi_{\Delta\leftrightarrow N\pi},
  I^\Delta_{\Delta\rightarrow N\pi} & \Delta \leftrightarrow N\pi \\[2mm]
  I^N_{N^*\rightarrow N\pi}, I^\pi_{N^*\leftrightarrow N\pi},
  I^{N^*}_{\Delta\rightarrow N\pi} & \Delta \leftrightarrow N\pi  \,\, ,
  \end{array}
\ee
denoting $N(1440)$ by $N^*$.

In the eqs.\  (\ref{BUUex}) all direct processes for pion production and
pion absorption are neglected. Since it is known that about 20\% of the pions
in nucleon--nucleon collisions, especially at low energies,  are produced in
direct processes, we have studied the effect of the direct channel
by  taking
into account also the
collision terms
\be
I^N_{NN\rightarrow NN\pi} \quad {\rm and} \quad I^N_{\pi NN\rightarrow NN}
\ee
for the change of the nucleon distribution function and the terms
\be
I^\pi_{NN\rightarrow NN\pi} \quad
{\rm and} \quad I^\pi_{\pi NN\rightarrow NN}
\ee
for the change of the pion distribution function due to collisions.

Since in  pion--nucleus reactions in the energy regime considered
single resonance excitations clearly dominate,
 we neglect resonance--resonance scattering which, however, is
included  for the nucleus--nucleus case \cite{ehe93i}.

\subsection{The test--particle method}

The standard way to solve the coupled nonlinear
integro--differential equations
is to discretize the distribution function, i.\ e.\ to substitute the
continuous distribution functions by a finite number of test--particles, i.\
e.\

\be
f(\vec r, \vec p;t) = {1\over N}
\sum_i^{N \cdot A}
\delta (\vec r- \vec r_i(t)) \delta (\vec p-\vec p_i(t))
\ee
where $N$ denotes the number of test-particles
per nucleon and A denotes the total
number of nucleons in the reaction. The ansatz (5) is a solution of the
Vlasov equation if $\vec r_i(t)$ and $\vec p_i(t)$ follow the classical
Hamilton-equations,
\bee
{d\vec p_i\over dt}& =& - \nabla _{\vec r} U\,\,  , \nonumber \\
{d\vec r_i\over dt}& =& {\vec p_i\over m_i}  \quad .
\eee

Thus the problem reduces to a classical  time
evolution of a system of a finite number of test-particles.
By discretizing the
time $t$ we solve this simplified problem as follows: For given
positions and momenta of the test-particles at time $t$ their evolution at
$t+\delta t$ reads:
\bee
\label{BUUp}
\vec p(t+\delta t)& =& \vec p(t) - \delta t\ \nabla _{\vec r}U(\vec r,t) \,\, ,
\\
\label{BUUr}
\vec r(t+\delta t) &= & \vec r(t) +\delta t \left(  {\vec p(t)\over E}
        \right)\,\, .
\eee

Instead of $\vec p/m$ we use $\vec p/E$ in the
Hamilton-equation as appropriate for
relativistic kinematics. We calculate the effect of collisions,
resonance decays
or absorptions between $t+\delta t/2$ and $t+3\ \delta t/2$ by
assuming straight line trajectories during this time interval.
For the simulation of inelastic reaction mechanisms we use the parallel
ensemble algorithm (ref. \cite{ber88}) in which collisions are only
 allowed between particles in the same ensemble. We note that the
isospin degrees  of freedom are accounted for by separate distribution
functions, but Coulomb--effects are not taken into account.

\subsection{Mean--field potentials}
\label{potentials}
The particles inside the nuclear medium propagate in different mean-field
 potentials as described
 in eqs.\ (\ref{BUUeq}),(\ref{BUUp}),(\ref{BUUr}). The nucleon mean-field
 potential is evaluated selfconsistently on the basis of a density--dependent
 Skyrme potential as:
\be
\label{nucpot}
U_N(\rho) = 0.75 t_0\rho + t_3\rho^{\tau}
\ee
with
$t_0=-1177.8MeV/fm^3$, $t_3=1845.5MeV/fm^5$ and $\tau = 5/3$, which leads
 to a binding energy of $-16MeV$ per nucleon at $\rho_0\approx 0.16 fm^{-3}$
and an
incompressibility  $K=308.4MeV$. These parameters give quite realistic, stable
density distributions for finite nuclei in the surface region.

The delta and $N(1440)$ resonance  are in general assumed to propagate
within the same mean field as the nucleons if not
explicitly noted otherwise.
Deviations from this
average potential will not only influence
the propagation of the resonances $R$
through the nucleus, but also change the collision probabilities; e.g. a
difference between the nucleon and the resonance potentials will shift
 the energy in the $RN\rightarrow NN$ reaction and the peak position in
the $\pi N\rightarrow R$ process. Thus to exploit the dynamical influence
 of an alternative  delta potential
we also incorporate different prescriptions for this mean field and discuss
the implications in comparison to the experimental data.

In our simulations we use two approximations for the delta potential.
The first model relates to elastic pion--scattering experiments
where one finds a delta potential
$V_\Delta \approx -30MeV$ at $\rho_0$ \cite{hor80}. Thus we have employed
  a delta potential which is just the  nucleon
potential shifted to $-30MeV$ at normal nuclear density,
\be
\label{shift}
U_{\Delta}(\rho) = 0.03 \frac{U_N(\rho )}{\mid U_N(\rho_0 )\mid} (GeV)\,\, .
\ee

As a second model  we used a delta potential as extracted by
Ehehalt et al.\ in \cite{ehe93} from the delta-hole model.
A suitable parametrization of this delta potential is given by \cite{ehe93}:
\be
\label{ehepot}
U_\Delta (\rho ) = -0.7 \rho + 1.75 \rho^{5/3}(GeV)  \quad .
\ee
This potential is also $-30MeV$ at $\rho_0$,
  but shows a different density
dependence.

Furthermore, in line with the delta-hole model also the pion propagates
in its own  potential field in the medium. We thus consequently also use
the respective pion potential as evaluated and parametrized in our
former work \cite{ehe93}, where we have investigated the efffect of
 pion and delta potentials
on asymptotic  pion spectra and  dilepton yields in heavy-ion collisons.

\subsection{Simulation of the collision integrals}

The test particles in our approach  collide with each other as in
conventional cascade simulations with collision rates as described
by  eq.\ (\ref{collproc}).
The reaction probabilities are calculated on the basis of free cross sections
for the different processes and implemented as
 explained in detail in \cite{wol90}  taking into account the
Pauli blocking of the nucleons in the final state with proper isospin factors.
The time--step used in the calculations is $\Delta t = 0.5 fm/c$;
we did not find any
significant change in the results when decreasing $\Delta t $ further.

\subsection{$\pi N$ cross section}

For the $\pi N \rightarrow \Delta$ cross section we employ  a Breit-Wigner
formula

\be
\label{pin}
\sigma_{\pi^+ p->\Delta^{++}} = \sigma^{max}_{\pi^+p}
\left(\frac{q_r}{q}\right)^2 \frac{\Gamma (M)}{(M_\Delta - M)^2+\frac{1}{4}
\Gamma (M)^2}
\ee

\noindent
with the  momentum-dependent width  \cite{koc84}:
\be
\Gamma (M) =  \left(\frac{q}{q_r}\right)^3 \frac{M_\Delta}{M} \left(
\frac{v(q)}{v(q_r)}\right)^2 \Gamma_r \,\, ,
\label{gamma}
\ee
where $M$ is the actual delta mass, $M_\Delta$ is the peak delta rest
mass ($1232MeV$), $q$ is the pion momentum in the
rest frame of the delta,  $q_r$ the value of $q$ for a delta  mass $M_\Delta$
and
$\Gamma_r = 110MeV$. The function
\be
v(q) =\frac{\beta^2}{\beta^2+q^2}
\ee
with $\beta =300MeV$ cuts the width at high momenta.

{}From the particle data table \cite{par90} we get
\bee
\sigma_{\pi^+p}^{max} &= & 200 mb \nonumber \\
\sigma_{\pi^-p}^{max} &= & 70 mb
\eee
and based on  isospin invariance of the strong interaction we use
\bee
\label{pimax}
\sigma_{\pi^-n}^{max} &= & 200 mb \nonumber \\
\sigma_{\pi^+n}^{max} &= & 70 mb  \\
\sigma_{\pi^0n}^{max}= \sigma_{\pi^0p}^{max} &= & 135mb \nonumber
\eee
in eq.\ (\ref{pin}) for the respective cross sections.

The production of the $N(1440)$ is only possible in the $P_{11}$ channel.
In order to   reproduce the $\pi^+n$ cross section around the
$N(1440)$ resonance we use the following form
of the cross section for the $\pi N \rightarrow N(1440)$ process
\be
\label{pinstar}
\sigma_{\pi^+n\rightarrow N^*} =\sigma_{\pi^-p\rightarrow N^*} =
\sigma_{\pi N\rightarrow N^*}^{max} \frac{0.25 \Gamma_{N^*}^2}
{(M_{N^*}-M)^2+\frac{1}{4}\Gamma_{N^*}^2}
\ee
with
$\sigma_{\pi N\rightarrow N^*}^{max}= 30mb$ and $\Gamma_{N^*} =200MeV$
which yields the cross section (dashed line) in fig.\ 1.
The sum of $\Delta$ and $N(1440)$ cross sections  (eq.\ (\ref{pin})
with eq.\ (\ref{pimax}) and eq.\ (\ref{pinstar})) then
 gives the correct value of the total cross section (cf. fig. 1
 (dashed--dotted  line)).

If a $\Delta$ or  $N(1440)$ is created in a $\pi N$
collision, its mass is determined by the
invariant mass  $\sqrt{s}$ of the $\pi N$ system; if it is created in a
$NN$--collision, then its mass is randomly chosen
weighted with the proper Breit--Wigner distribution (see eq.\
(\ref{massdist})).
In our approach the resonances
are treated as on--shell particles, however, with dynamically
determined mass and
lifetime (see below).

\subsection{Decay of resonances}
\label{Ddecay}
The decay of a resonance $R \rightarrow \pi N$ in free space
 is determined by its life time $\Gamma(M)^{-1}$.
However, inside the nucleus the resonance decay may be
forbidden by the Pauli--principle. In addition, a resonance
can also decay by the process
 $R N \rightarrow N N$ due to the presence of other nucleons. We will
discuss the latter channel in the next subsection and describe here
the numerical implementation of the 'free' decay
inside the nucleus.

Both resonances, $\Delta$ and $N(1440)$, decay with $99.5\%$ and $65\%$,
respectively,  into
a $\pi N$ final state. Thus in every time--step we calculate the decay
probability into a $\pi N$ state in the rest frame of the resonance
by assuming an exponential decay law

\be
\label{propdecay}
P_{decay}=\rme^{-\Gamma (M)\Delta t}
\ee
where $\Gamma (M)$ is the width of the resonance and $\Delta t$ is the
time--step
in our calculation. By a Monte Carlo method we then
decide, if the resonance actually decays in a given time--step.
In case of a $\Delta$ we use eq.\ (\ref{gamma}) for the width, for
the $N(1440)$
a constant width $\Gamma = 200 MeV$.
Before we let the resonance decay we check if the outgoing nucleon is
Pauli blocked by counting the number of
testparticles in the final phase space cell as described in \cite{ber88}.

The $\Delta$ decay, furthermore, is assumed to be isotropic in its rest frame
and the pion and nucleon momenta
are determined by energy and momentum conservation.
Nevertheless, to test
the influence of an anisotropic delta decay we also have performed
 calculations
with a p--wave angular distribution
for the differential $\pi N$ cross section
\be
\label{uniso}
\frac{d\sigma}{d\Omega} \sim (1+3\cos ^2\Theta )
\ee
where $\Theta$ is the angle
between the initial and final pion in the delta rest frame (c.f. section
\ref{ineldif}).

\subsection{Direct pion emission and absorption}
\label{direct}

Whereas in the VerWest--Arndt parametrization  for the
$NN\rightarrow N\Delta\rightarrow NN\pi$ process (\cite{ver82} c.f.
 section \ref{DNNN})
all pions are assumed to be produced via a delta  or a
$N(1440)$--resonance up to $\sqrt{s}\approx 2.5 GeV$, the pion may
also be created directly in a NN collision. We calculate the
direct  pion production cross section, which is the production
via an off--shell nucleon,
in lowest order perturbation theory using  a
One--Boson--Exchange model for the
nucleon--nucleon interaction. The full theory, that
takes into account intermediate deltas and nucleons,
reproduces the experimental data rather well \cite{sch93}.
 To extract the direct contribution to the production process
we  dropped all diagramms containing  a
delta in the intermediate state. (For a
more detailed description of the calculations we refer the reader to
ref.\ \cite{sch93}.)

For the different isospin channels we use the isospin decompositon of
the $NN\rightarrow NN\pi$ process (according to  ref.\ \cite{ver82}):
\be
\label{decomp}
\begin{array}{lc}
pp\rightarrow pp\pi^0 & \sigma_{11} \\
pp\rightarrow pn\pi^+ & \sigma_{11} +\sigma_{10}\\
np\rightarrow np\pi^0+ & \frac{1}{2}(\sigma_{10} +\sigma_{01})\\
np\rightarrow nn\pi^++ & \frac{1}{2}(\sigma_{11} +\sigma_{01})\\
np\rightarrow pp\pi^-+ & \frac{1}{2}(\sigma_{10} +\sigma_{01})
\end{array}
\ee
where $I,I'$ in $\sigma_{I,I'}$ denotes the initial and final isospin
 of the participating nucleon pair. By calculating the direct
contribution to the $pp\rightarrow pn\pi^+$, $pp\rightarrow pp\pi^0$ and
$np\rightarrow np\pi^0$ processes we determine all isospin channels using
decomposition eq. (\ref{decomp}). Our result for the direct
$pp\rightarrow np\pi^+$
process is shown in fig. 2 (solid line). It
 has the following functional dependence on $\sqrt{s}$ (in $GeV$):
\be
\sigma_{pp\rightarrow np\pi^+}^s(\sqrt{s})= 26.2-40.6712\sqrt{s}
+13.55566s         \quad .
\ee
For the other isospin channels we used the same energy dependence with the
following  ratios (calculated at $\sqrt s=2.155 GeV$):
\be
\sigma_{pp\rightarrow np\pi^+}:\sigma_{np\rightarrow np\pi^0}
:\sigma_{pp\rightarrow pp\pi^0} = 6.98:2.269:1\quad .
\ee

Taking into account the direct pion production process, we simultaneously
have to consider the inverse process $\pi NN\rightarrow NN$. We avoid
the problem
of calculating a cross section with a three--particle initial state
 by using  the quasi 'deuteron' assumption; the pions are assumed
to be absorbed on nucleon pairs as in the conventional
 picture for
pion absorption \cite{wey90}. The $\pi (2N)\rightarrow NN$ reaction we
then can calculate from
the direct production cross sections by using detailed balance:
\be
\label{pidirect}
\sigma_{\pi NN\rightarrow NN}=\frac{4}{3}\frac{p_f^2}{p_i^2}
\sigma^s_{NN\rightarrow NN\pi}
\ee
with
\bee
\label{pii}
p_i^2&=&(s-(m_\pi-2m_n)^2)(s-(m_\pi+2m_n)^2)/(4s)\\
p_f^2&=& \frac{s}{4}-m_n^2  \quad ,
\eee
which denote the initial $(2N)$ pair and
final nucleon momenta in the center of
mass frame.
In fig. 2 we show the $\pi^+(pn)\rightarrow pp$ cross section  used
in the calculation by the dashed line.
The isospin dependence discussed above
leads to the following absorption
 ratios:
\bee
\frac{\sigma_{\pi^+(nn)\rightarrow np}}
{\sigma_{\pi^+(np)\rightarrow pp}}  &=&  0.083 \nonumber\\
\frac{\sigma_{\pi^0(np)\rightarrow np}}
{\sigma_{\pi^+(np)\rightarrow pp}}  &=&  0.44 \nonumber\\
\frac{\sigma_{\pi^0(nn)\rightarrow nn}}
{\sigma_{\pi^+(np)\rightarrow pp}}  &=&  0.14 \\
\frac{\sigma_{\pi^0(pp)\rightarrow pp}}
{\sigma_{\pi^+(np)\rightarrow pp}}  &=&  0.14 \nonumber\\
\frac{\sigma_{\pi^-(pp)\rightarrow np}}
{\sigma_{\pi^+(np)\rightarrow pn}}  &=&  0.083 \nonumber\\
\frac{\sigma_{\pi^-(np)\rightarrow nn}}
{\sigma_{\pi^+(np)\rightarrow pp}}  &=&  1 \nonumber \quad .
\eee

\subsection{$N\Delta\leftrightarrow NN$ and $NN(1440)
\leftrightarrow NN$ reactions}
\label{DNNN}

For the $NN\rightarrow N\Delta$ and $NN\rightarrow NN(1440)$ processes
 we use the VerWest--Arndt parametrization
of the cross section as described in detail in ref.\ \cite{wol90}.
The
mass of the resonance $R$, which is populated in our simulation
in a $NN$--collison, is chosen according to the probability function
\be
\label{massdist}
F(M^2) = \frac{1}{\pi} \frac{M_\Delta \Gamma (M)}
{(M^2-M_\Delta^2)^2 + M_\Delta^2\Gamma (M)^2}
\ee
 with $\Gamma (M)$ for deltas given in eq.\ (\ref{gamma}) and $\Gamma (M)
 =200MeV$ for  $N(1440)$.
  In a
pion--nucleus reaction
these resonance production
processes are negligible, but the inverse processes
 $N\Delta\rightarrow NN$ and $N(1440)N\rightarrow NN$
are very important in the context of pion absorption.
In the literature the latter reactions are determined from the
measured cross sections for the inverse channels by detailed balance
using slightly different assumptions \cite{Wol93,wol92,ber88,dan91},
that have to be discussed in more detail. The ambiguity arises because
the measured cross sections given in  ref.\ \cite{ver82} are
averaged over the $\Delta$--mass distribution.

Application of detailed balance to a hypothetical delta with fixed
mass  leads to \cite{ber88}
\be
\label{DB}
\sigma_{\Delta^{++}n\rightarrow pp} = \frac{1}{4}
\frac{p_f^2}{p_i^2}\sigma_{pp\rightarrow n\Delta^{++}}
\ee
for the $\Delta^{++}n\rightarrow pp$  cross section. In eq.\ (\ref{DB})
 $p_f,p_i$ are the final and initial momenta of the particles in the
center of mass frame whereas the factor
$1/4$ is due to  spin averaging and a symmetry factor
for identical particles in
the final state.

Taking into account, however,  that the delta has no fixed
mass in our simulation, we get\cite{wol92}:
\be
\label{deltdb}
 \sigma_{n\Delta^{++}\rightarrow pp}^{tot}  = \frac{2\pi}{4}
\frac{p_N^2}{p_\Delta^2}
\frac{1}{F(M^2)}
\int_{-1}^{1}\frac{d \sigma_{pp\rightarrow n\Delta^{++}}}{d(\cos{\Theta})
 dM^2} d(\cos \Theta ) \quad .
\ee
Since the  VerWest--Arndt parametrization only gives the $\Delta$--mass
averaged total pion production
cross section it cannot be used to calculate the inverse
cross section in eq.\ (\ref{deltdb}). Following our former work \cite{wol92}
we  therefore use
\be
\label{gyu}
 \sigma_{n\Delta^{++}\rightarrow pp}= \frac{4 \pi}{4}
\frac{p_N^2}{p_\Delta^2}
 \sigma_{pp\rightarrow n\Delta^{++}}
\frac{1}
{\int_{(m_N+m_\pi)^2}^{(\sqrt{s}-m_N)^2} f(M^2)dM^2}\,\,
\ee
with  the $NN\rightarrow N\Delta$ cross sections from VerWest--Arndt.

To clarify the situation concerning the different expressions  in
the literature
for the $N\Delta\rightarrow NN$ reaction we inserted in eq.\ (\ref{deltdb})
the result for the mass-dependent cross
 section $d\sigma_{pp\rightarrow n\Delta^{++}}/(d\Omega dM^2)$
from a calculation of Sch\"afer et al.\ \cite{sch93}.
The latter differential cross section for the $pp\rightarrow n\Delta^{++}$
process is calculated on the basis of a OBE model using pion-- and
 $\rho$--exchange which reproduces the experimental data for
 $d\sigma / dM$ (cf. fig. 3) remarkable well.

To investigate the validity of our model assumption in \cite{wol92} we
compare the results of eqs.\ (\ref{DB}),(\ref{gyu}) and eq.\
(\ref{deltdb}) calculated with
the OBE model for the $\Delta^{++}n\rightarrow pp$ cross section for
different delta masses $M=1180, 1232$ and $1280MeV$ in fig. 4. Our
 cross section  (solid line) eq.\ (\ref{gyu})
is in  good agreement with the OBE result (dotted line)
for larger invariant energy
 $\sqrt{s}$, but  overestimates the absorption cross section close to
threshold.  In fig.\ 4  also the cross section as proposed
 by Danielewicz and Bertsch in ref.\  \cite{dan91}
is shown by the dashed--dotted line. It fits better
 at small $\sqrt{s}$ but overestimates the absorption significantly
at higher energy
$\sqrt{s}$. The naive detailed balance (\ref{DB}) (dashed line)
underestimates the cross section at almost all energies.
For the dynamical implications of the different models in
pion--nucleus reactions we refer the reader to
sections \ref{abs} and \ref{inel}.

Finally, using isospin factors and taking into account symmetry factors for
identical particles in the final state,
we obtain  the total cross sections   of the other
isospin channels as
\bee
\sigma_{n\Delta^{+}\rightarrow pn}^{tot}&  = &
\sigma_{p\Delta^{0}\rightarrow pn}^{tot}  =
\frac{2}{3} \sigma_{n\Delta^{++}\rightarrow pp}^{tot} \nonumber \\
\sigma_{p\Delta^{+}\rightarrow pp}^{tot}&  = &
\sigma_{n\Delta^{0}\rightarrow nn}^{tot}  =
\frac{1}{3} \sigma_{n\Delta^{++}\rightarrow pp}^{tot} \nonumber \\
\sigma_{p\Delta^{-}\rightarrow nn}^{tot}&  = &
 \sigma_{n\Delta^{++}\rightarrow pp}^{tot} \quad .
\eee

In case of  the $NN(1440)\rightarrow NN$ process we use eq.\ (\ref{gyu})
with a constant width $\Gamma_{N^*}=200MeV$ and the $NN\rightarrow
NN(1440)$ cross section from the VerWest--Arndt parametrization.
Additionally, we also check for Pauli blocking of the outgoing nucleons;
however,  due  to the large momenta transfered to both nucleons the
blocking effect is found to be of minor importance in pion--nucleus
reactions.

\subsection{$NN$, $N\Delta$ and $N(1440)N$ elastic scattering}

For these processes we use the conventional Cugnon
 parameterization \cite{cug82,ber88}
\be
\sigma (mb) = \frac{35}{1+100 \sqrt{s}\, '(GeV)}
\ee
with $\sqrt{s}\, '= \sqrt{s} - 1.8993$ for a $NN$ collision and
$\sqrt{s}\, '=\sqrt{s}-0.938-M_R$ for nucleon-resonance elastic scattering.
The anisotropic angular distribution used  is adopted from \cite{wol90}.

\subsection{Inclusive cross sections}
\label{crosssection}
To calculate  inclusive cross sections for a pion-nucleus reaction
we perform an explicit impact parameter
integration,
\be
\sigma = \int 2\pi bdb\,\,\, N_{reac}(b) \quad ,
\ee
where $N_{reac}(b)$ denote the impact--parameter dependent
particle multiplicities as obtained from
simulations with fixed $b$ for the pions.

This technique is identical to the one employed for proton-nucleus or
nucleus-nucleus reactions so that we describe all reactions on the same
footing. In ref.\ \cite{wol90,ehe93} we have already
shown that the experimental
spectra on pion production in heavy--ion collisions can be reproduced
very well, except for some small differences at small pion transverse
momenta. If this also holds for pion-nucleus collisions will be studied in
the next section.

\section{RESULTS FOR PION-NUCLEUS REACTIONS}

\subsection{Total absorption cross sections}
\label{abs}
As noted in the introduction the mechanism of pion
 absorption in nuclei is not well understood.
Energy and momentum conservation rule out absorption of a pion on
a free single nucleon. However, due
 to the large momentum mismatch in the reaction the single--nucleon
 absorption is very unprobable  in pion--nucleus reactions, too.
 The most important mechanism is thus the
two--body absorption, where the pion is absorbed by a pair of nucleons.
The quantitative contribution of higher order processes and especially
that of   three--body processes is still a matter of debate.
For heavier nuclei the three--body process is controversely discussed to
contribute
between 10\%
and 50\% of the total absorption cross section \cite{ing93}.

In the transport simulations the pion absorption mainly proceeds
via a two--step mechanism: a pion first is captured in a $\Delta$ state and
this $\Delta$ can be  absorbed in a second step via the channel
$\Delta N\rightarrow NN$. This two--step process clearly simulates a two--body
absorption mechanism. However,
by allowing the $\Delta$ to decay to $\pi N$ and the outgoing pion
 to be captured  again in another
$\Delta$ (subsequent two-body reaction chain), we incorporate also higher
order absorption mechanisms with on--shell
intermediate pions.

In fig.\ 5 we show the calculated mass dependence of the pion
 absorption cross section
at the pion energy $T_\pi = 165MeV$ in comparison to the experimental data.
 To illustrate the influence of
the $\Delta N\rightarrow NN$ reaction we have performed
 calculations with the different detailed balance prescriptions
 discussed in section \ref{DNNN}. We find that our previously
used formula eq.\ (\ref{gyu}) \cite{wol92}(solid line), the OBE results eq.\
(\ref{deltdb}) (dotted line) and the prescription from ref.\
\cite{dan91}(dashed--dotted line)
 lead to the same accuracy in  reproducing
the experimental data, whereas the naive detailed balance (eq.\ (\ref{DB}))
misses about 50\% of the cross section. This reflects the fact, that
the  detailed balance prescription of eq.\ (\ref{DB}) underestimates
the elementary absorption cross section over the whole
energy regime (see fig. 4) and explains why the cascade calculation of
\cite{cug88} underestimated the absorption cross section by a factor of 2.

Turning now to the energy dependence  of the absorption
cross section (fig.\ 6),
we find that we can describe the cross section at low energies up
to the resonance region very well for all target masses.
For higher energies ($T_\pi=315 MeV$), however,  our calculation (solid line)
overestimates the data. This disagreement becomes larger for  heavier
nuclei. It is worthwhile to note that even in this
$T_\pi$ regime the $A$--dependence of the absorption cross section is well
described. The
energy dependence of the pion absorption  is not
altered  when using the OBE model (dotted line)
or the prescription from ref.\ \cite{dan91}(dashed--dotted line)
(see section \ref{DNNN}) for the $\Delta N\rightarrow NN$
reaction.

Compared to other calculations for the total absorption
cross section as a function of target mass and kinetic energy of the pion
\cite{fra82,sal88} our results are of comparable quality.
The INC calculations have problems in reproducing
the mass dependence of the total absorption cross section \cite{fra82}
and also  the absolute value in the resonance region,
whereas the calculations of Salcedo et al.\ \cite{sal88} also overestimate
pion absorption at the higher pion energies for heavier nuclei.

To simulate  the effect  of
a direct three--body absorption process in our approach, we used  a
density dependent modification for the
$\Delta N \rightarrow NN$ cross section
of the form:
\be
\label{DNrho}
\sigma'_{\Delta N\rightarrow NN} = (1 +3\frac{\rho}{\rho_0}) \sigma_{\Delta
N \rightarrow NN}   \quad ,
\ee
 thus incorporating the density dependence of  three--body pion absorption
in an ad--hoc way.
The result for the energy--dependent absorption cross section on
$^{12}C$ and $^{209}Bi$ with the
 density--dependent cross section  (\ref{DNrho})
is shown in fig.\ 7 (dashed line).
We  find only a minor three--body effect for the $^{209}Bi$ case.
This is  due to the
large diameter  of the $Bi$--nucleus;
most deltas inside the $Bi$--nucleus scatter several times with
nucleons and are absorbed.
An increase of the  $\Delta N\rightarrow NN$ cross section as introduced
in eq.\ (\ref{DNrho}) just
leads to an earlier absorption and not to an increase of the total
pion absorption cross section.

For the $^{12}C$ nucleus we find a bigger effect of the three--body
absorption process as
simulated by prescription (\ref{DNrho}); fig.\ 7 (dashed
line). In this lighter nucleus  deltas do not scatter as often as in
the $Bi$--case and
 can leave the nucleus without being
absorbed. Since the escape probability of the $\Delta$'s
increases with their velocity, modifications
of eq.\ (\ref{DNrho}) for the
$\Delta N\rightarrow NN$ process affect the total
pion absorption cross section  dominantly  at higher energies,
thus overestimating the total pion absorption
cross section
 even more.

Salcedo et al.\ \cite{sal88} report a larger contribution of the three--body
absorption processes
of about 30\%, but these calculations  would lead to similar
results as ours, if they omitted their three--body absorption probabilities.
Similar to our
case without three--body absorption probabilities the pions
 in their calculation would be absorbed further
inside the nucleus via a two--body process. We thus conclude that
the total inclusive  pion absorption process in heavy nuclei
is not
 sensitive to true many--body effects.
It seems that subsequent
two body steps, taking into account off--shell deltas, can
simulate the many--body effects.

We additionally note, that we have also studied
the influence of the $N(1440)$ for the absorption
process and have found only a 10\% difference
for $T_\pi\approx 300MeV$ when performing  the
calculations  with and without the
$N(1440)$ resonance. For smaller energies the contribution
of the  $N(1440)$ is negligible.

As mentionend in section \ref{direct} we studied also
 the effect of direct (NN) pion
production and absorption. We find a 10\% enhancement for the absorption
cross section at all pion energies in the $^{12}C$ case,
 whereas in the $^{209}Bi$ case the pions are absorbed earlier
in the reaction due to the incorporation of the direct
process, but the total absorption cross section
is not changed (see discussion on many--body absorption above).

Summarizing this section we may state that we reproduce the experimental
data on pion absorption quite well, for light nuclei at all
energies and for heavy nuclei for energies up to the
$\Delta$--resonance region; for heavy nuclei above the $\Delta$--resonance
we obtain a too large cross section. The origin of
this deficiency may be in the cascade description itself. We find
that for central ($\pi ,A$) collisions the absorption cross section even has
a minimum at $T_\pi=160MeV$ because at this resonance energy the
$\Delta$--excitation cross section is very large and the delta can thus be
 formed already  in the nuclear surface, where
 the probability for
the decay pion to escape the nucleus again (inelastic scattering) is very
large. The calculated energy dependence of the absorption cross section,
which shows a different behaviour, is thus largely
determined by the impact parameter dependence.

\subsection{Total inelastic cross section}
\label{inel}

Inelastic pion-nucleus scattering occurs in competition with
the absorption process. Thus to understand the pion--nucleus
reaction both processes must be described within the same
approach.

As shown in fig.\ 8  our calculations reproduce  the mass dependence
 of the total inelastic cross section
 for pion energies ($T_\pi = 165 MeV$) (solid line).
 This also holds when employing  the OBE model for
 the $\Delta N\rightarrow NN$ channel (dotted line) or the
prescription by Danielewicz \cite{dan91} (dashed--dotted line).

When looking at the energy dependence of the
 inelastic cross sections, which is shown
in fig.\ 9 for $^{12}C$ and $^{209}Bi$, we
 see that for the  $^{209}Bi$ case we underestimate the cross section
by about a factor 2 at all pion energies; for
$^{12}C$ this  only happens for higher energies above resonance.

A detailed phase-space analysis offers the following kinematic reason
for these deficiencies:
Generally most of the pions that are scattered inelastically emerge
from the first $\Delta$ generation which is dominantly created in the
surface of the nucleus; approximately 50\% of the pions from the first
$\Delta$ generation leave the system again and are not absorbed.
When increasing the pion kinetic energy above the resonance region
the cross section for delta
formation becomes smaller and consequently the
first delta generation is created further inside the nucleus, such
that pions from its decay cannot
leave the nucleus as easily as in the resonance region.
Furthermore, these  secondary pions are shifted down in energy (see section
\ref{deltdyn}),
which leads to an effective increase of the reabsorption cross section.
As a consequence the secondary pions in
the interior of the nucleus are almost
completely absorbed, which leads to a reduced inelastic pion cross section and
a correspondingly enhanced absorption cross section for
higher incident pion energies.

At low energies, on the other hand,
the first generation of deltas
has  a
larger life time due to their  low mass
(see section \ref{deltdyn}).
This does not change the picture just described in
heavy nuclei, but in light nuclei some of these
deltas can thus  decay outside the nucleus
and the secondary pions can escape again.

\subsection{Differential inelastic cross section}
\label{ineldif}
Since in the resonance region the inelastic cross
sections are well reproduced,
we proceed with calculating also differential cross sections to get further
information on the  initial delta distributions.
In fig.\ 10 the results of our simulations for ${\pi^+}+^{12}C$
 and ${\pi^+}+^{209}Bi$
at $T_\pi =160MeV$ are shown for the differential inelastic cross section
 (solid line). Except for the forward angles the  data
for  $^{209}Bi$ are rather well  reproduced both with
 an isotropic and an anisotropic delta decay
(see discussion in section \ref{Ddecay}).
For the $^{12}C$ target, however,  the
 anisotropic delta decay gives clearly a
better description of the data at backward angles
 (dashed line in fig.\ 10).
The deficiency  in  forward direction is due to the fact that
the  forward scattered pions will be captured in a delta again so that
it becomes rather improbable to leave the nucleus in forward direction.
Furthermore, the forward amplitude is
also expected to be sensitive to coherent quantum scattering \cite{sal88};
this effect is completely missing in our incoherent transport approach.

Using the more realistic anisotropic delta decay instead of the
isotropic one we find that this effects the  total cross section for absorption
and
inelastic scattering by at most 5\% such  that the basic conclusions of the
previous sections remain unchanged.
Again our results are comparable  with those of Salcedo
et al. (ref.\ \cite{sal88}) and are quantitatively better than
the results from INC calculations \cite{fra82}.

\subsection{$(\pi^+,Np)$ reactions}

A further test for the absorption mechanism in our model is the comparison to
more exclusive
experiments where also the continuum protons are detected in a pion
 absorption event. In a recent
study  Ransome et al.\ \cite{ran92} present
total cross sections for pion absorption
on nuclei followed by 2p,3p emission for
different energies and masses.
We have calculated the $(\pi^+,2p)$ and $(\pi^+,3p)$
cross sections for $T_\pi=150 MeV$, where our calculation
reproduces the total absorption
cross section. For the continuum protons
we applied the same  cut in energy $E_p > 23 MeV$
as in the experiment. To calculate the cross section
we proceeded as described in section \ref{crosssection}. For
different impact parameters we counted the absorption
events followed by two, three proton emission, respectively.

Our calculated results are compared in table \ref{tab1}
 with the
data; they are in good agreement, especially for the lighter nuclei.
For the heavier nuclei we get too many $(\pi^+,3p)$ events, but for this
channel the errors in the experimental data are bigger. In accordance
with the  experimental data
we find that
the contribution of $(\pi^+,Np)$ processes with $N>3$ to
 the absorption is negligible.

Thus we conclude that the average event characteristics are well described
within our transport approach.

\begin{table}
\begin{tabular}{c||c|c||c|c}
{\rm element} & $(\pi^+,2p\&3p)$ &
$(\pi^+,2p\&3p)$ & $(\pi^+,3p)$ & $ (\pi^+,3p)$ \\
&{\rm BUU}   &{\rm exp.}  & {\rm BUU}  &  {\rm exp.}   \\
\hline \hline
$^{12}C $       & 114         &  110       &  15        &   14        \\
$^{58}Ni $      & 272        & 300        &  30        &   33         \\
$^{118}Sn $     & 333        & 320        &  46        &   31         \\
$^{208}Pb $     & 399        & 400        &  48        &   24
\end{tabular}
\caption{\label{tab1} Cross sections for $(\pi^+,Np)$ reactions in $mb$
for $T_\pi=150MeV$
 from ref.\ [39].
We compare with the sum of $2p$ and  $3p$ events, because these are
the  experimental data given under the heading total 2p. 
}
\end{table}

\subsection{Pion and delta dynamics}
\label{deltdyn}

In the calculations described so far the deltas have been propagated
in the same potential as the nucleons.
Recently \cite{cug88,ehe93,sne93}, there has been some
discussion of the properties of the delta
in the nuclear medium. In order to explore the physical
consequences of any in--medium
changes we
have performed calculations changing artificially the peak mass
and width of the resonance.
To be more precise, in
the width $\Gamma_{\Delta}(M)$  we have changed the parameter $\Gamma_r$
 in eq.\ (\ref{gamma}), which consequently
influences all other cross sections  that
 depend on  $\Gamma(M)$. In varying systematically the peak mass of the delta
and its width at maximum we find that both changes do not improve the
energy--dependence of the absorption cross section
discussed earlier in sections
\ref{abs} and \ref{inel}. The best reproduction of the absorption
data is achieved for the free width and the free delta peak mass
(solid line in fig.\ 11).
Thus our model analysis of pion absorption cross sections
indicates no need for any in--medium changes of the $\Delta$--resonance.

In a recent publication Sneppen and Gaarde
\cite{sne93} have argued that the analysis of charge exchange
reactions in terms of a cascade model requires
an in--medium delta mass of $M_\Delta =1202MeV$ and a width of
$\Gamma (M_\Delta) = 200MeV$.
In a calculation with these parameters we find
for the $^{12}C$ target (fig.\ 12, (dashed line))
that the agreement with the experimental
data at higher energies is better compared  to the original
 calculation  (solid line),
 but becomes worse for low $T_{\pi}$.
In addition,  the inelastic scattering cross section at higher energies
 is not well described. For the $^{209}Bi$ case (not
 explicitly displayed) there is not much change.

Another in-medium effect for the deltas is the delta potential as
already pointed out in section \ref{potentials}. In order to explore
the sensitivity of the pion-nucleus data to pion and delta selfenergies
 we have studied the influence of pion and delta potentials given in
section \ref{potentials} in our simulations. Whereas the effect of a
 pion potential changes the result only on a 5\% level (not explicitly
displayed), the pion absorption shows a larger
sensitivity to a $\Delta$--potential as shown in fig.\ 13.
We find that the effective potential from the delta-hole model
eq.\ \ref{ehepot} (dotted line)
and the simple shifted potential eq.\ (\ref{shift}) (dashed line)
lead to similar results. The calculated changes
in the absorption cross section
can be directly traced back to the energy shift
 in the $\pi N$ reaction.

Furthermore, to separate kinematical effects in the
 delta and pion dynamics in the pion--nucleus
reactions we have calculated the differential $\Delta$ mass distribution
 for different $\Delta$ generations.
The results in fig.\ 14 suggest the following picture: At
 higher energies also high mass deltas
are excited, as naively expected. These deltas in turn have a short life time
(large $\Gamma(M)$) and decay fast. The pions from their
decay are reabsorbed again and regenerate the $\Delta$'s
although now at a smaller mass due to recoil effects.
Thereby  the $\Delta$'s cascade
down in mass and loose about $70MeV$ in mass per generation.

\section{SUMMARY}

We have analysed  pion--nucleus reactions within the extended BUU model
that had originally been developed for the description of nucleus-nucleus
collisions. Due to the pionic entrance channel we expected to be
 especially sensitive  to the pion and delta dynamics employed and
 thus to be able to put a stringent constraint on these ingredients
for the description of heavy--ion collisions.

Apart from heavy--ion collisions we  can reproduce the inclusive data
for pion-nucleus reactions for a wide range of pion kinetic energies
and target  masses. Within the transport approach we have simulated
pion absorption by sequential binary reactions between pions, nucleons
and delta's. The measured pion absorption cross sections
are very well reproduced for pion energies up to the
resonance energy; for higher energies we overestimate
the
absorption. The mass dependence, on the other side, is very well
reproduced.  The dominant part of the many--body absorption effects
seems to be simulated by the dynamics
incorporated in our transport approach.
The deficiencies of the model could essentially
be traced back to the first
generation of deltas.

Apart from pion absorption we also reproduce
the inelastic cross sections for light nuclei
up to the resonance region  quite well. For heavy nuclei we
seem to underestimate the inelastic cross section at all energies.

We obtain  a good agreement for differential inelastic scattering
 and the exclusive $(\pi^+,Np)$ reactions in
the delta resonance region which demonstrates that the overall event
pattern is described quite well in this region.

In summary we find that the BUU transport approach
developed for nucleus-nucleus
collisions provides a rather reliable description of pion-nucleus reactions,
too. The results obtained for the different channels
in the pion--nucleus reactions are in their quality comparable to those of
other recent model
calculations for these reactions.

\newpage
\section*{Figure Captions}

\bigskip
\noindent
{\bf Fig. 1:} Decomposition of the $\pi^-p$ cross section:
Breit--Wigner cross section
for $\pi^-p\rightarrow \Delta^0$ (eq.\ (\ref{pin}))
(solid line) and Breit--Wigner cross
section for $\pi^-p\rightarrow N(1440)$ (eq.\ (\ref{pinstar}))
(dashed line). Sum of eq.\ (\ref{pin}) and eq.\ (\ref{pinstar})
(dashed--dotted line). Experimental
data are from ref.\ \cite{bal88}.

\bigskip
\noindent
{\bf Fig. 2:} Cross sections for the
direct $pp\rightarrow pn\pi^+$ process
 based on the OBE model (see text) (solid line)
and the direct $\pi^+(pn)\rightarrow pp$ process (\ref{pidirect})
(dashed line)  as a function of
$\sqrt{s}$.

\bigskip
\noindent
{\bf Fig. 3:} The mass dependent cross section for
$pp\rightarrow \Delta^{++}n$ at $\sqrt{s}=1.48$ in
the OBE model from Sch\"afer et al. \cite{sch93}. Experimental data
are from ref.\ \cite{dim86}.

\bigskip
\noindent
{\bf Fig. 4:} The cross section for
$\Delta^{++}n\rightarrow pp$ within different
descriptions for the detailed balance as a function
 on $\sqrt{s}-M$:
naive detailed balance eq.\ (\ref{DB})
(dashed line), calculation with the
OBE model eq.\ (\ref{deltdb}) (dotted line),
eq.\ (\ref{gyu})  (solid line),
 Danielewicz  \cite{dan91} (dashed--dotted line).
The calculations are
for delta masses $M=1180,1232$ and $1280 MeV$.

\bigskip
\noindent
{\bf Fig. 5:} Mass dependent total absorption cross section for $\pi^+$
at $T_\pi = 165MeV$ within  different descriptions
for the detailed balance:
naive detailed balance of eq.\ (\ref{DB})
(dashed line), calculation with the
OBE model eq.\ (\ref{deltdb}) (dotted line),
eq.\ (\ref{gyu}) (solid line),
 Danielewicz  \cite{dan91} (dashed--dotted line).
Experimental data are from ref.\ \cite{ash81}.

\bigskip
\noindent
{\bf Fig. 6:} Total absorption cross section for $\pi^+$ on
$^{12}C$, $^{56}Fe$ and $^{209}Bi$ as a function of the
pion energy:
Calculation with the OBE model eq.\ (\ref{deltdb}) (dotted line),
eq.\ (\ref{gyu}) (solid line),
 Danielewicz  \cite{dan91} (dashed--dotted line).
 Experimental data are from
ref.\ \cite{ash81}.

\bigskip
\noindent
{\bf Fig. 7:} Total absorption cross section for $\pi^+$ on
$^{12}C$ and $^{209}Bi$ as function of the pion energy:
Calculation with
eq.\ (\ref{gyu}) (solid line) and eq.\ (\ref{DNrho}) for
the $\Delta N \rightarrow NN$ process (dashed line).
Experimental data are from
ref.\ \cite{ash81}.

\bigskip
\noindent
{\bf Fig. 8:} Mass dependent total inelastic cross section for $\pi^+$
at $T_\pi = 165MeV$:
Calculation with the
OBE model eq.\ (\ref{deltdb}) (dotted line),
eq.\ (\ref{gyu}) (solid line),
 Danielewicz  \cite{dan91} (dashed--dotted line).
Experimental data are from \cite{ash81}.

\bigskip
\noindent
{\bf Fig. 9:} Energy dependence of the  inelastic
 absoroption cross section for $\pi^+$ on
$^{12}C$ and  $^{209}Bi$:
Calculation with the OBE model eq.\ (\ref{deltdb}) (dotted line),
eq.\ (\ref{gyu}) (solid line),
 Danielewicz  \cite{dan91} (dashed--dotted line).
 Experimental data are from
ref.\ \cite{ash81}.

\bigskip
\noindent
{\bf Fig. 10:} Differential inelastic cross section for $\pi^+$ on $^{209}Bi$
(upper figure) and on $^{12}C$ (lower figure)
at $T_\pi =160MeV$. Calculation
with isotropic delta decay (solid line) and with anisotropic decay eq.\
(\ref{uniso}) (dashed line). Experimental data are from ref.\ \cite{lev83}.

\bigskip
\noindent
{\bf Fig. 11:} Total absorption cross section for $\pi^+$ on
$^{12}C$ and $^{209}Bi$ as a function of the pion energy.
Calculations with different delta rest masses are shown in the
left figure and with different $\Gamma_r$ in eq.\ (\ref{gamma}) for
all cross sections in the right  figure.
 Experimental data are from
ref.\ \cite{ash81}.

\bigskip
\noindent
{\bf Fig. 12:} Total absorption (upper figure) and
total inelastic (lower figure) cross section for $\pi^+$ on
$^{12}C$ as a function of the pion energy.
Calculations with the parameters
from ref.\ \cite{sne93}: delta rest mass $M_\Delta = 1202 MeV$ and
$\Gamma_r=200MeV$ only in $\Gamma (M)$  of eq.\ (\ref{propdecay}).
 Experimental data are from
ref.\ \cite{ash81}.

\bigskip
\noindent
{\bf Fig. 13:} Energy dependence of the  total
absorption cross section for $\pi^+$ on
$^{12}C$ and $^{209}Bi$. Calculations with different delta potentials:
$V_\Delta =V_N$ eq.\ (\ref{nucpot}) (solid
line), shifted delta potential eq.\ (\ref{shift}) (dashed line)
and potential from ref.\
\cite{ehe93} eq.\ (\ref{ehepot})(dotted line). Experimental data are from
ref.\ \cite{ash81}.

\bigskip
\noindent
{\bf Fig. 14:} Delta mass distribution for different delta generations for
$\pi^+$ on $^{209}Bi$ at $85MeV$ (lower figure) and at $245MeV$
(upper figure). Free mass distribution eq.\ (\ref{massdist}) (dashed line),
first delta generation (solid line), second delta
generation (dashed--dotted line), third delta generation (dotted line) and
fourth delta generation (thick dashed line).

\end{document}